# Si$_{1-x-y}$Ge$_y$Sn$_x$ alloy formation by Sn ion implantation and flash lamp annealing

O. Steuer[1,4], M. Michailow[2], R. Hübner[1], K. Pyszniak[3], M. Turek[3], U. Kentsch[1], F. Ganss[1], M. M. Khan[1], L. Rebohle[1], S. Zhou[1], J. Knoch[2], M. Helm[1,5], G. Cuniberti[4], Y. M. Georgiev[1,6], S. Prucnal[1]

[1] Institute of Ion Beam Physics and Materials Research, Helmholtz-Zentrum Dresden-Rossendorf, Bautzner Landstrasse 400, 01328 Dresden, Germany

[2] Institute of Semiconductor Electronics, Rheinisch-Westfälische Technische Hochschule Aachen, 52074 Aachen, Germany

[3] Institute of Physics, Maria Curie-Sklodowska University in Lublin, M. Curie-Sklodowskiej 1, 20-031 Lublin, Poland

[4] Institute of Materials Science and Max Bergmann Center, Technische Universität Dresden, 01069 Dresden, Germany

[5] Center for Advancing Electronics Dresden, Technische Universität Dresden, 01062 Dresden, Germany

[6] Institute of Electronics, Bulgarian Academy of Sciences, 72, Tsarigradsko Chausse Blvd.,1784 Sofia, Bulgaria

E-mail: o.steuer@hzdr.de, s.prucnal@hzdr.de

# Abstract

For many years, $Si_{1-y}Ge_y$ alloys have been applied in the semiconductor industry due to the ability to adjust the performance of Si-based nanoelectronic devices. Following this alloying approach of group-IV semiconductors, adding tin (Sn) into the alloy appears as the obvious next step, which leads to additional possibilities for tailoring the material properties. Adding Sn enables effective band gap and strain engineering and can improve the carrier mobilities, which makes $Si_{1-x-y}Ge_ySn_x$ alloys promising candidates for future opto- and nanoelectronics applications. The bottom-up approach for epitaxial growth of $Si_{1-x-y}Ge_ySn_x$, e.g., by chemical vapor deposition and molecular beam epitaxy, allows tuning the material properties in the growth direction only; the realization of local material modifications to generate lateral heterostructures with such a bottom-up approach is extremely elaborate, since it would require the use of lithography, etching, and either selective epitaxy or epitaxy and chemical-mechanical polishing giving rise to interface issues, non-planar substrates, etc. This article shows the possibility of fabricating $Si_{1-x-y}Ge_ySn_x$ alloys by Sn ion beam implantation into $Si_{1-y}Ge_y$ layers followed by millisecond-range flash lamp annealing (FLA). The materials are investigated by Rutherford backscattering spectrometry, micro-Raman spectroscopy, X-ray diffraction, and transmission electron microscopy. The fabrication approach was adapted to ultra-thin $Si_{1-y}Ge_y$ layers on silicon-on-insulator substrates. The results show the fabrication of single-crystalline $Si_{1-x-y}Ge_ySn_x$ with up to 2.3 at.% incorporated Sn without any indication of Sn segregation after recrystallization via FLA. Finally, we exhibit the possibility of implanting Sn locally in ultra-thin $Si_{1-y}Ge_y$ films by masking unstructured regions on the chip. Thus demonstrating the realization of vertical as well as lateral $Si_{1-x-y}Ge_ySn_x$ heterostructures by Sn ion implantation and flash lamp annealing.

# I Introduction

The incorporation of tin (Sn) in silicon-germanium (SiGe) can reduce the band gap and enhance the carrier mobilities in the resulting $Si_{1-x-y}Ge_ySn_x$ alloys [1, 2]. This material modification is not causing difficulties, since Sn does not lead to problematic contamination, and therefore, ternary $Si_{1-x-y}Ge_ySn_x$ alloys are compatible with complementary metal-oxide-semiconductor (CMOS) technology. Currently, the main fabrication methods for these metastable Sn-containing alloys are epitaxial low-temperature growth by molecular beam epitaxy (MBE) [3, 4] and chemical vapor deposition (CVD) [5, 6]. The vertical growth enables the precise adjustment of the alloy composition in the growth direction but does not allow lateral modifications. On the other hand, ion beam implantation followed by non-equilibrium thermal treatments can accomplish lateral and even local material modifications. Unfortunately, high-fluence ion beam implantation into $Si_{1-y}Ge_y$ alloys comes with some challenges, like i) maintaining a smooth surface during ion implantation and ii) overcoming thermal limitations in the recrystallization process if the concentration of the implanted species exceeds the equilibrium solubility. Especially high-fluence implantation in Ge-based alloys causes strong surface degradation [7, 8]. This issue can be overcome by inserting capping layers and reducing the temperature during the implantation [9-11]. The recrystallization after Sn implantation requires non-equilibirum thermal treatments, since the solid solubility of Sn in $Si_{1-y}Ge_y$ is less than 1 at.% and decreases with increasing Si concentration [12]. Nanosecond-range pulsed laser annealing (PLA) and millisecond-range flash lamp annealing (FLA) have already been applied to recrystallize implanted layers [10, 11, 13-16]. The PLA approach seems to have limitations towards higher Sn concentrations. Tran et al. reported vertical filaments (similar to those shown after post-growth pulse laser melting of $Ge_{1-x}Sn_x$ [17, 18]) after PLA of $Ge_{1-x}Sn_x$ (x = 4.8, 5.25 and 6 at.%) at λ = 355 nm for 6 ns [15] and significant Sn redistributions in the $Ge_{0.91}Sn_{0.09}$ layer [9]. Additionally, the intermixing of oxygen (O) from the $SiO_2$ cap, due to recoil implantation with $Ge_{1-x}Sn_x$, can cause the formation of oxide clusters after PLA [9]. On the other hand, FLA of $Ge_{0.955}Sn_{0.045}$ from the rear-side (r-FLA) at $E_d$ = 65 J cm$^{-2}$ for 3.2 ms enabled the recrystallization by solid-phase epitaxy without Sn redistribution [11, 19]. About 95% of Sn atoms were located at Ge substitutional sites. To the best of our knowledge, no successful fabrication of $Si_{1-x-y}Ge_ySn_x$ by Sn ion beam implantation followed by FLA has been reported so far. Using commercially available silicon-on-insulator (SOI) substrates, growing $Si_{1-y}Ge_y$ thin films by state-of-the-art CVD and fabricating $Si_{1-x-y}Ge_ySn_x$ by Sn implantation and r-FLA is another important step towards the ability to realize lateral $Si_{1-x-y}Ge_ySn_x$ devices. However, the thin films and the isolating $SiO_2$ require precise tuning of the process parameters. In this paper, we show the general feasibility of fabricating $Si_{1-x-y}Ge_ySn_x$ by Sn implantation and r-FLA, transfer this alloying approach to ultra-thin $Si_{1-y}Ge_y$ films on SOI substrates, and present the concept of local Sn implantation into $Si_{1-y}Ge_y$ on SOI substrates.

## II Experimental part

Three different $Si_{1-y}Ge_y$ materials were implanted with Sn, schematically shown in Fig. 1 a) and b). Sample type A has a 600 nm-thick single-crystalline $Si_{0.28}Ge_{0.72}$ top layer, epitaxially grown on 675 µm-thick Si (001) substrates by MBE. Sample type B has a 15 nm-thick single-crystalline $Si_{0.73}Ge_{0.27}$ layer grown by CVD on a commercial ultra-thin SOI substrate. The SOI wafer itself consists, from top to bottom, of a 12 nm-thick top Si (100) layer, a 21 nm-thick buried oxide (BOX) film, and a 775 µm-thick Si carrier wafer. Sample type C has a 10 nm-thick $Si_{0.70}Ge_{0.30}$ layer covered with a 3 nm-thick Si layer and is grown on an SOI substrate. This SOI substrate has a 90 nm-thick top-Si and a 145 nm-thick BOX layer. Before Sn implantation, the surfaces of sample types B and C were covered with an 8.6 nm-thick SiN$_x$ and a 10 nm-thick SiO$_2$ capping layer, respectively. Additionally, sample type C was covered with a 115 nm-thick amorphous Si (a-Si) film, and an implantation window was opened by using a combined spacer and damascene process followed by selective wet-chemical etching, as described in supplemental materials A. The Sn implantation parameters were selected according to "stopping and range of ions in matter" simulations (SRIM-code) [20] and are presented in supplemental materials B. Sample type A was implanted with $^{120}$Sn ions with an acceleration voltage of $E$ = 250 kV and a fluence $D$ = 1 x 10$^{16}$ cm$^{-2}$. Sample type B was implanted with $^{120}$Sn ions at $E$ = 26 kV and $D$ = 1.2 x 10$^{15}$ cm$^{-2}$. Sample type C was $^{120}$Sn - implanted with the beam parallel to the cavity opening (7°), or 14° tilted relative to the cavity opening, at an acceleration voltage of 24 kV and a fluence $D$ = 1x 10$^{15}$ cm$^{-2}$. For all implantations, the sample holder was actively cooled with liquid nitrogen from the back side (cold finger), and an implantation tilt of 7° relative to the crystal surface was used. The implantation tilt helps to reduce ion channelling effects during the implantation process.

Post-implantation annealing was performed for sample types A and B by r-FLA in nitrogen (N$_2$) atmosphere, with pre-heating, a flash pulse length of 3.2 ms or 20 ms, and energy densities of 72.5 J cm$^{-2}$ or 150 J cm$^{-2}$, respectively. The FLA system, schematically shown in Fig. 1 c), is equipped with Xe flash lamps, and halogen lamps below the sample are used for pre-heating. The energy density on the sample surface is separately measured with a power probe (Fit for Intense Pulse Light (Fit-IPL-R)) of the company "Laserpoint". The r-FLA pre-heating temperature corresponds to the maximum temperature at the top surface of a 525 μm-thick Si wafer, which was placed between the sample front side and the underlying quartz window.

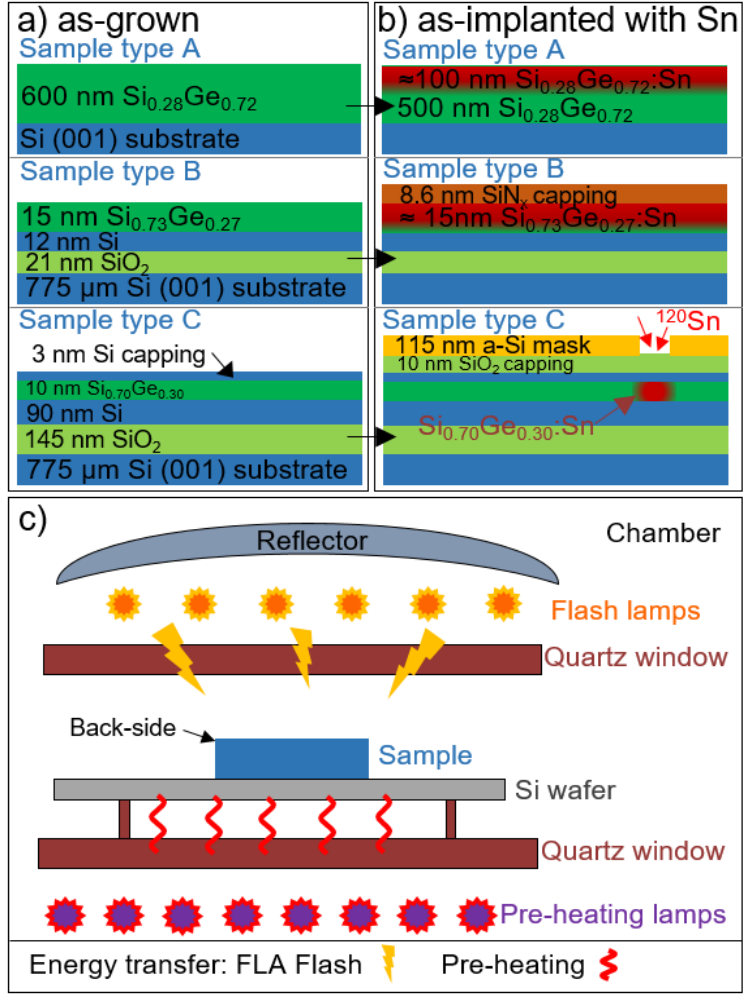

Fig. 1: Layer stacks of the investigated sample types A, B, and C in the as-grown state a) and after $^{120}$Sn ion implantation b). The FLA tool is schematically illustrated in c), and the samples are placed upside down on the Si wafer to perform r-FLA.

Samples were characterized in the as-grown state, after Sn ion implantation, and after r-FLA by μ-Raman spectroscopy, Rutherford backscattering spectrometry (RBS), X-ray diffraction (XRD), and cross-sectional transmission electron microscopy (TEM).

RBS random (RBS-R) and channeling (RBS-C) measurements were performed using a 2 MV Van-de-Graaff accelerator. The used He$^+$ beam had an energy of 1.7 MeV, beam currents between 10 and 20 nA, and was confined by an aperture with a diameter of 1 mm. Each measurement was performed with a detector angle of 170°. The obtained RBS spectra were fitted with the software SIMNRA [21] to calculate the Si$_{1-x-y}$Ge$_y$Sn$_x$ layer thickness and alloy composition. RBS-C was performed along the [001] crystal axis.

The RBS channeling yield $\chi$ was calculated by Eq. 1, where $A_C$ is the integrated area under the channeling curve and $A_R$ is the integrated area under the random curve.

$$\chi = \frac{A_C}{A_R} \qquad \text{Eq. 1}$$

Furthermore, the incorporation rates of Sn on Ge lattice sites $\xi_{Sn,Ge}$, and Sn on Si lattice sites $\xi_{Sn,Si}$ in the same layer depth were calculated using Eq. 2.

$$\xi_{Sn,Ge} = \frac{(1-\chi_{Sn})}{(1-\chi_{Ge})} \text{ and } \xi_{Sn,Si} = \frac{(1-\chi_{Sn})}{(1-\chi_{Si})} \qquad \text{Eq. 2}$$

µ-Raman spectroscopy was performed on a Horiba LabRam n°1/24 h system equipped with a neodymium-doped yttrium aluminum garnet (Nd:YAG) laser with a wavelength of 532 nm. All measurements were performed at ambient conditions in backscattering geometry with a circular laser diameter of 1 µm. The scattered light was diffracted by an 1800 lines / mm grating and finally detected by a liquid-nitrogen-cooled charge-coupled device (CCD). XRD was performed on a Rigaku SmartLab X-ray diffractometer system equipped with a copper X-ray source and a Ge (220) two-bounce monochromator. High-resolution XRD (HR-XRD) *θ-2θ* scans were carried out on the symmetrical 0 0 4 reflections. The reciprocal space maps (RSM) were generated for the asymmetrical 2 2 4 reflections by scanning *ω* and measuring the diffracted intensity in dependence of 2 *θ* with the detector in 1D single-exposure mode. The alignment for HR-XRD and RSM was performed on the 0 0 4 and 2 2 4 Si substrate reflections, respectively. Cross-sectional bright-field and high-resolution TEM was performed using an image-C$_s$-corrected Titan 80-300 microscope (FEI) operated at an accelerating voltage of 300 kV. High-angle annular dark-field scanning transmission electron microscopy (HAADF-STEM) imaging and spectrum imaging analysis based on energy-dispersive X-ray spectroscopy (EDXS) were performed at 200 kV with a Talos F200X microscope equipped with a Super-X EDX detector system (FEI). Prior to (S)TEM analysis, the specimen mounted in a high-visibility low-background holder was placed for 8 s into a Model 1020 Plasma Cleaner (Fischione) to remove potential contamination.

## III. Results and discussion

### A) $Si_{1-x-y}Ge_ySn_x$ thick film on Si substrate

The capability of the $Si_{1-x-y}Ge_ySn_x$ alloying approach by Sn ion beam implantation into $Si_{1-y}Ge_y$ layers followed by r-FLA recrystallization was probed with relatively thick films of sample type A. The µ-Raman and RBS results of the as-grown state, after implantation, and after r-FLA are shown in Fig. 2. In the as-grown state, well-visible Ge-Ge and Si-Ge phonon modes are obtained at 295.6 and 405.5 ± 0.1cm$^{-1}$. After Sn implantation, the Ge-Ge and Si-Ge modes disappear, since the selected fluence exceeds the amorphization threshold. r-FLA recrystallizes the layer effectively. Hence, the Ge-Ge and Si-Ge phonon modes occur at 294.4 ± 0.1cm$^{-1}$ and 401.0 ± 0.1cm$^{-1}$. Furthermore, both phonon modes are shifted to lower wavenumbers after r-FLA. The Ge-Ge shift could be caused by increasing the in-plane lattice parameter due to in-plane tensile strain or incorporating the larger Sn atoms into the $Si_{1-y}Ge_y$ lattice. On the other hand, the Si-Ge mode position depends additionally on the ratio between Si and Ge in the $Si_{1-y}Ge_y$ alloy. Sn incorporation is suggested, since the formation of significant tensile strain is unlikely and both modes shift in the same direction.

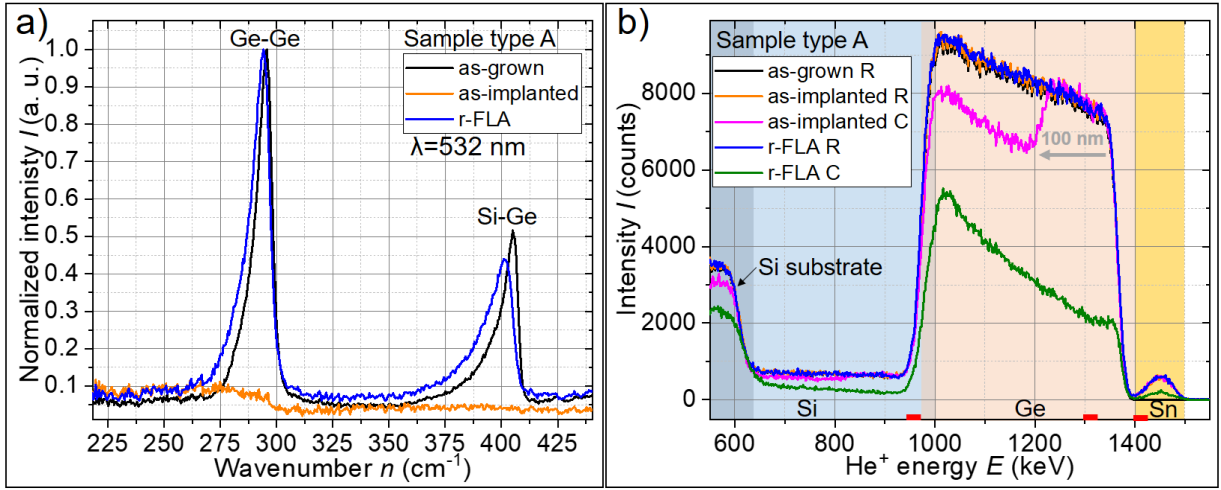

Fig. 2: a) µ-Raman and b) RBS-R/C measurement results of the $Si_{1-x-y}Ge_ySn_x$ alloy (sample type A) in the as-grown state before implantation, after low-temperature Sn implantation, and after alloy formation via r-FLA with a pre-heating step at 330 °C for 30 s and a flash length of 3.2 ms with $E_d$ = 72.5 J cm$^{-2}$. The RBS spectra contain contributions of Sn (1400 - 1500 keV, orange background), Ge (930 - 1400 keV, beige background), and Si (up to 1000 keV, blue background). The RBS-C alignment was carried out along the [001] crystal axis. The red-marked He$^+$ energy ranges in b) belong to the same layer depth of the $Si_{1-x-y}Ge_ySn_x$ alloy used for the RBS-R/C analysis.

RBS-R/C measurements in Fig. 2 b) were performed to confirm the incorporation of Sn into the $Si_{0.28}Ge_{0.72}$ lattice. After the implantation, an Sn signal occurred in the RBS-R/C spectra between 1400 and 1500 keV. Furthermore, the absence of channeling in the Sn contribution and in the first about 100 nm of the Ge contribution confirms the amorphization of the entire implanted layer. The RBS-R signal after r-FLA is almost identical to the as-implanted state. Hence, significant diffusion or redistribution of Sn could be suppressed due to the short annealing time. A total Sn concentration of about 2.3 at.% was determined by SIMNRA fitting. The analysis of the r-FLA spectra using Eq. 1 and Eq. 2 reveals an Sn channeling yield $\chi_{Sn}$ of 31% and an almost full incorporation of Sn on Ge ($\xi_{Sn,Ge}$ = 95%) and Si ($\xi_{Sn,Si}$ = 100%) sites, respectively. The incorporation of Sn is also in line with the interpretation of the Raman spectra, where the Ge-Ge and Si-Ge phonon modes shifted to lower wavenumbers after r-FLA (see Fig. 2 a). Therefore, we conclude that ion beam implantation of Sn into $Si_{1-y}Ge_y$ followed by FLA can be used to fabricate $Si_{1-x-y}Ge_ySn_x$ alloys.

B) $Si_{1-x-y}Ge_ySn_x$ thin film on SOI substrates

The gained process know-how was transferred to the much more challenging layer stack of sample type B. The significantly higher Si concentration of $Si_{0.73}Ge_{0.27}$ reduces the maximum Sn solid solubility in the alloy [12]. On the other hand, the stability of $Si_{1-y}Ge_y$ against surface degradation might be higher, since Si [22] is more resistant to surface modifications caused by ions than pure Ge [7, 8]. Furthermore, the SOI substrate reduces heat conduction during the ion implantation and r-FLA, and the thin layers require precise process parameters.

An overview of the three different fabrication states (as-grown, as-implanted, and recrystallized by r-FLA) can be obtained from the cross-sectional TEM results presented in Fig. 3. TEM-based images of the as-grown state in Fig. 3 a) - c) show an almost homogeneous high-quality $Si_{0.73}Ge_{0.27}$ layer on the SOI substrate. After the implantation, quite regularly recurring amorphous regions with wave-like shape appear along the single-crystalline $Si_{1-y-x}Ge_ySn_x$ layer. More details about the amorphous structures are presented in Supplement C. A high-resolution image of such an amorphized region is presented in Fig. 3 d), and the interface between the amorphous and crystalline structures is highlighted in Fig. 3 e). The origin of these amorphous structures might be the interplay between the selected dose, current density, implantation depth, heat transfer, sample cooling efficiency, and alloy composition, which caused locally different amorphization or *in situ* recrystallization conditions. To the best of our knowledge, such amorphous structures have not yet been reported in the literature. On the other hand, the amorphous structures do not influence the sample surface roughness. The corresponding EDXS analysis results in Fig. 3 f) and g) show a homogeneous Sn distribution within the $Si_{1-y-x}Ge_ySn_x$ layer. The stronger Sn signal at the sample surface might be caused by Sn migration from the $SiN_x$ layer during the EDXS measurement.

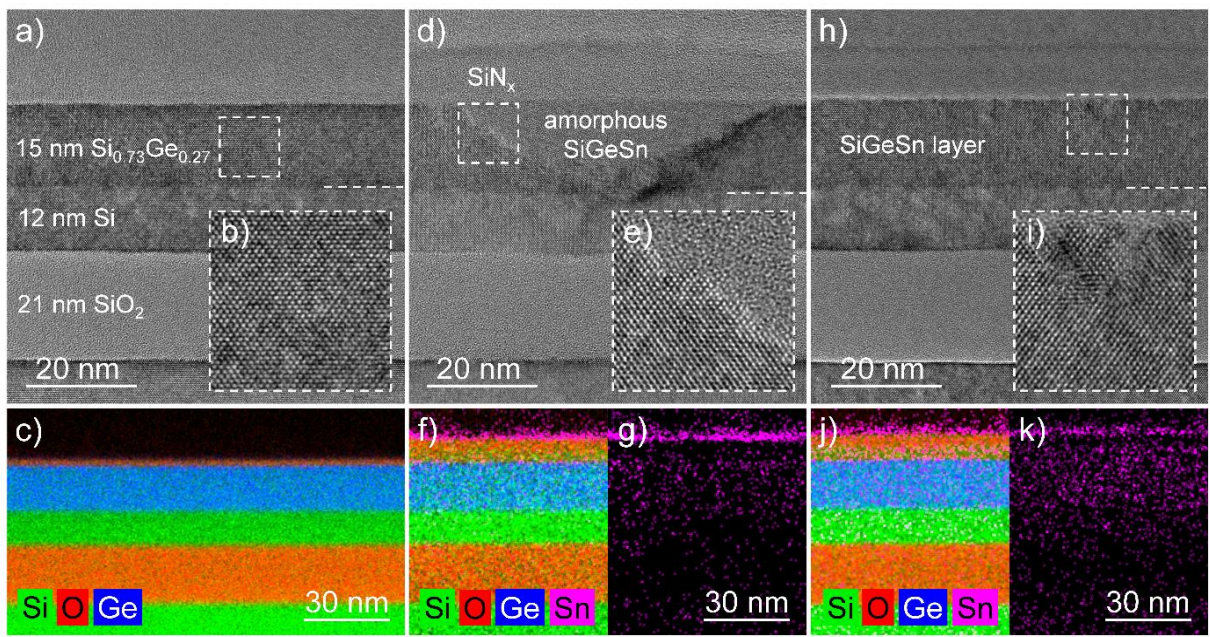

Fig. 3: High-resolution TEM images (a), b), d), e), h), and i)) and EDXS-based element distribution maps of $Si_{1-x-y}Ge_ySn_x$ on SOI (sample type B) in the as-grown state (a) - c)), after Sn implantation (d) - g)), and after r-FLA with a peak pre-heating temperature of 570 °C and flash processing at $E_d$ = 150 J cm$^{-2}$ for 20 ms (h) - k)). The $Si_{1-x-y}Ge_ySn_x$ / SOI interface is highlighted with a white-dashed line in a), d), and h). The insets b), e), and i) are enlargements of the outlined regions in the corresponding images a), d), and h), respectively. Panels c), f), and j) show superimposed EDXS-based element distributions with silicon (Si) in green, oxygen (O) in red, germanium (Ge) in blue, and tin (Sn) in magenta. Panels g) and h) show only the respective Sn signal.

After r-FLA, the $Si_{1-y-x}Ge_ySn_x$ layer is completely recrystallized, as visible in Fig. 3 h). Only a few lattice defects are observed (Fig. 3 i)). According to Fig. 3 j) and k), the Sn distribution is homogeneous and comparable to the as-implanted state shown in Fig. 3 g). In particular, there are no indications for Sn segregation within the $Si_{1-y-x}Ge_ySn_x$ layer.

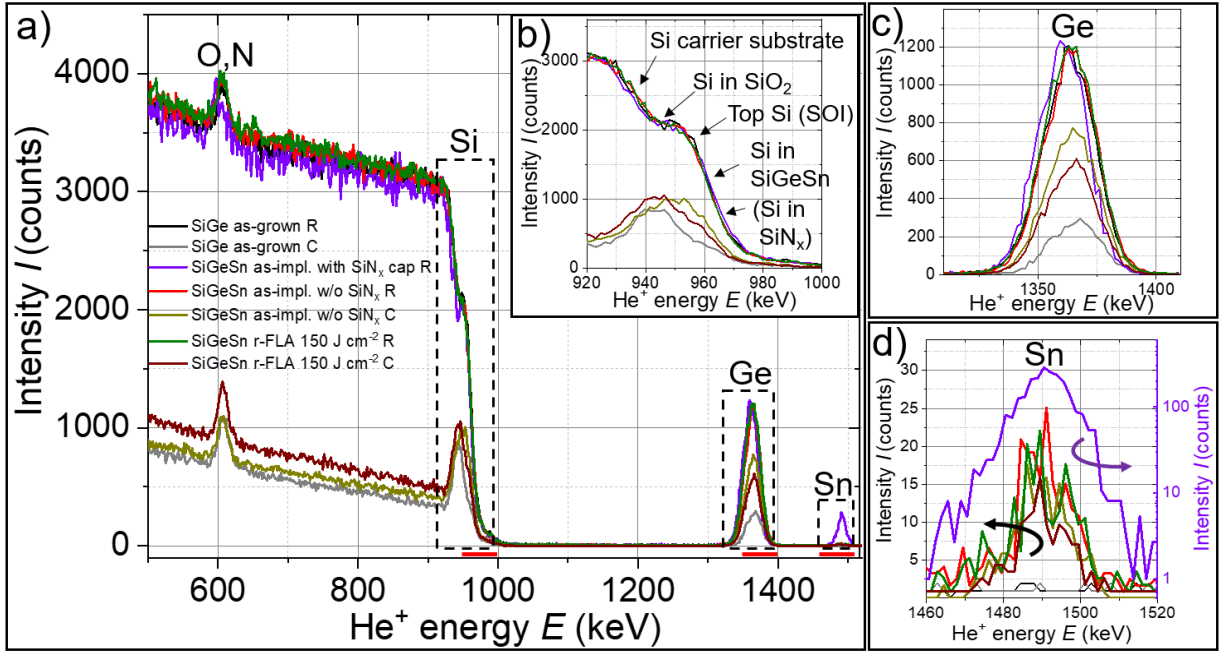

Fig. 4: RBS-R/C results of the $Si_{1-x-y}Ge_ySn_x$ alloy in the as-grown state before implantation, after Sn implantation (with and without $SiN_x$ cap), and after r-FLA with peak pre-heating at 570 °C and an energy density of 150 J cm$^{-2}$ for 20 ms a). Enlargements of the spectra regions for Si b), Ge c), and Sn d).

The RBS-R/C measurements presented in Fig. 4 were performed to simulate the implanted Sn concentration and to evaluate the Sn incorporation in the lattice. The RBS spectra contain contributions from Sn (1460 – 1520 keV), Ge (1330 – 1400 keV), Si (200 – 1000 keV), and O / N (580 – 620 keV), as allocated in Fig. 4 a). The broad Si contribution comes from the entire layer stack (see Fig. 4 b)) and is a mixture of the contributions from i) Si in $SiN_x$ at the surface ("SiGeSn as-implanted sample with $SiN_x$ cap"), ii) Si in the $Si_{1-x-y}Ge_ySn_x$ alloy, iii) Si of the top SOI layer, iv) Si in the buried oxide (BOX) $SiO_2$ layer, and v) Si of the bulk Si substrate. For a detailed comparison of the $Si_{1-x-y}Ge_ySn_x$ layer properties, the channeling yield $\chi$ and the occupation of Sn on Ge or Si sites were calculated by Eq. 1 and Eq. 2 and are presented in Table 1. The $Si_{0.73}Ge_{0.27}$ as-grown RBS-R/C spectra show pronounced channeling of Si ($\chi_{Si}$ = 27.2%) and Ge ($\chi_{Ge}$ = 25.7%) in Fig. 4 b) and c) and confirm the epitaxial growth of the single-crystalline $Si_{0.73}Ge_{0.27}$ layer on SOI. After implantation, a significant Sn signal occurred in the as-implanted state with the $SiN_x$ capping layer. This confirms the implantation of Sn, but the presence of $SiN_x$ causes significant dechanneling. After removing the $SiN_x$ layer by etching with 40% HF:DI for 1 min, the Sn intensity is reduced (see Fig. 4 d)). This indicates that a significant amount of Sn was implanted in the $SiN_x$ cap, which coincides with the EDXS results in Fig. 3 g). Poor channeling ($\chi_{Si} \approx$ 49%, $\chi_{Ge} \approx$ 63%, and $\chi_{Sn} \approx$ 71%) could be measured in the as-implanted state due to the *in situ* recrystallization during the implantation process. After r-FLA and SiN$_x$ removal, the Ge and Sn peak shapes are similar to those in the as-implanted state. Furthermore, the channeling yield of Si ($\chi_{Si} \approx 40\%$), Ge ($\chi_{Ge} \approx 47\%$), and Sn ($\chi_{Ge} \approx 60\%$) decreased due to the recrystallization of the amorphous structures and thermally activated healing of displaced atoms. In general, $\chi$ is much higher than expected from the TEM results in Fig. 3 i). This is related to an increased surface roughness after SiN$_x$ removal, which causes dechanneling. However, the incorporation rates of Sn after r-FLA indicate an increased incorporation of Sn on the Si sites.

Table 1: RBS-R/C analysis results of the as-grown Si$_{0.73}$Ge$_{0.27}$ reference, as well as the Si$_{1-x-y}$Ge$_y$Sn$_x$ alloy in the as-implanted state and after r-FLA. The minimum channeling yield for Si $\chi_{Si}$, Ge $\chi_{Ge}$, and Sn $\chi_{Sn}$, as well as the substitutional fraction of Sn on substitutional Si sites $\xi_{Sn,Si}$ and on substitutional Ge sites $\xi_{Sn,Ge}$, were calculated by Eq. 1 and Eq. 2. Integration intervals between 950 and 1000 keV for Si, 1340 and 1390 keV for Ge, and 1465 and 1515 keV for Sn were used.

| Sample | $\chi_{Si}$ (%) | $\chi_{Ge}$ (%) | $\chi_{Sn}$ (%) | $\xi_{Sn,Si}$ (%) | $\xi_{Sn,Ge}$ (%) |
|---|---|---|---|---|---|
| Si$_{0.73}$Ge$_{0.27}$ as-grown | 27.2 | 25.7 | - | - | - |
| SiGe:Sn as-implanted w/o SiN$_x$ | 48.9 | 63.3 | 70.6 | 57.5 | 80.1 |
| Si$_{1-x-y}$Ge$_y$Sn$_x$ r-FLA 150 J cm$^{-2}$ 20 ms | 39.6 | 47.3 | 60.4 | 65.5 | 75.1 |

In order to determine the Si$_{1-x-y}$Ge$_y$Sn$_x$ alloy composition, additional RBS-R measurements were performed with an incident angle of 80° for samples after SiN$_x$ removal. The SIMNRA simulation results reveal a chemical composition of approximately Si$_{0.681}$Ge$_{0.308}$Sn$_{0.011}$.

Since the incorporation of Sn is relatively small, XRD $\theta$-$2\theta$ scans at the 0 0 4 reflection and RSMs around the 2 2 4 reflection of Si were performed as a complementary investigation method to RBS. The HR-XRD scans in Fig. 5 contain a broad Si$_{1-x-y}$Ge$_y$Sn$_x$ 0 0 4 reflection between 67° and 68.5° and a sharp Si 0 0 4 reflection of the substrate and the top SOI layer at 69.15°, which is close to the literature value of pure Si at 69.13° [23].

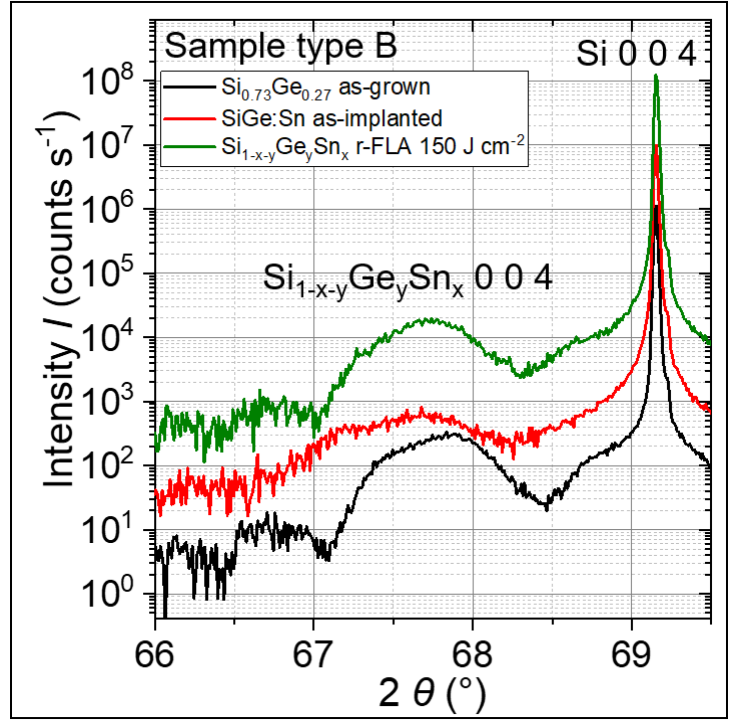

Fig. 5: 0 0 4 HR-XRD results of the $Si_{1-x-y}Ge_ySn_x$ alloy in the as-grown state, after implantation and after r-FLA with a peak pre-heating temperature of 570 °C and r-FLA parameters of $E_d$ = 150 J cm$^{-2}$ and a pulse length of 20 ms. For better visibility, the curves gained an offset in intensity.

As-grown $Si_{0.73}Ge_{0.27}$ has a relatively broad $Si_{1-y}Ge_ySn_x$ reflection, which is related to a small chemical gradient across the $Si_{0.73}Ge_{0.27}$ layer and the general small layer thickness. After implantation, the intensity of the $Si_{1-x-y}Ge_ySn_x$ 0 0 4 reflection is reduced, since the number of coherent lattice planes is reduced due to the local amorphization (see Fig. 3 c)). After r-FLA at $E_d$ = 150 J cm$^{-2}$ for 20 ms, the $Si_{1-x-y}Ge_ySn_x$ layer is fully recrystallized. Thus, the intensity of the $Si_{1-x-y}Ge_ySn_x$ 0 0 4 reflection increased again. Additionally, a peak extension towards smaller diffraction angles is observed due to the incorporation of Sn in the lattice. Another feature in the HR-XRD scans are Laue oscillations [24] of the $Si_{1-x-y}Ge_ySn_x$ 0 0 4 reflection in the as-grown state and after r-FLA at $E_d$ = 150 J cm$^{-2}$ for 20 ms. Visible fringes indicate a high layer crystal quality and a sharp interface between the layers [24].

Fig. 6 shows the 2 2 4 RSMs of the investigated samples. In the as-grown state, the $Si_{0.73}Ge_{0.27}$ 2 2 4 reflection has the same in-plane lattice parameter as the Si substrate. Hence, Fig. 6 a) confirms a fully pseudomorphic growth of the SiGe layer on the SOI wafer. On the other hand, the implantation damage in combination with partial *in situ* recrystallization leads to the first relaxation events in the $Si_{1-x-y}Ge_ySn_x$ layer, as shown in Fig. 6 b). After r-FLA at $E_d$ = 150 J cm$^{-2}$ for 20 ms (see Fig. 6 c)), the $q_x$ and $q_z$ distributions are almost the same as in the as-implanted state, but the overall intensity is increased due to recrystallization.

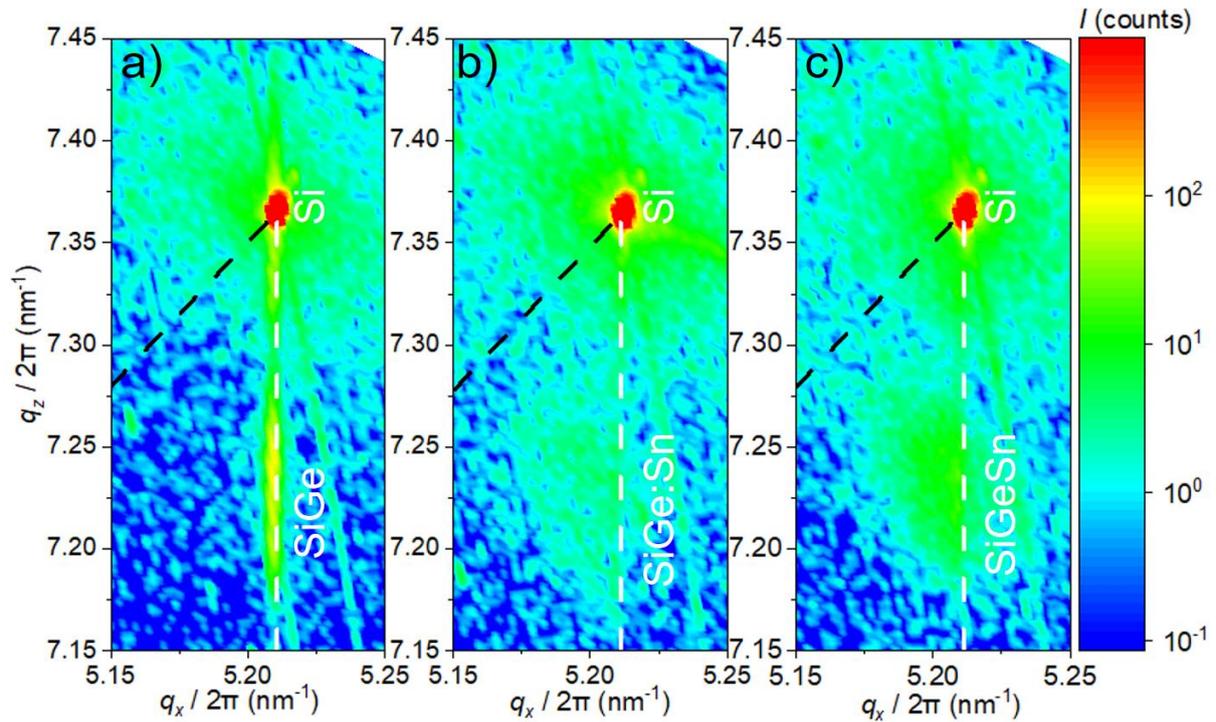

Fig. 6: 2 2 4 XRD-RSM of the $Si_{0.73}Ge_{0.27}$ as-grown state a), the as-implanted SiGe:Sn state b), and $Si_{1-x-y}Ge_ySn_x$ after pre-heating at 570 °C and r-FLA with $E_d$ = 150 J cm$^{-2}$ for 20 ms c) from sample type B. The white vertical dashed line corresponds to the fully pseudomorphically grown state, and the black dashed line is the strain relaxation line for a fully relaxed alloy on Si.

### C) Local lateral Sn implantation

To showcase the feasibility of local Sn implantation, sample type C was structured and Sn-implanted at 14° out-of-axis (see Fig. 7 a), d)) and parallel to the cavity (see Fig. 7 b), c), e) and f)). Owing to the high-fluence Sn implantation at low temperatures, the former single-crystalline $Si_{1-y}Ge_y$ converted into an amorphous $Si_{1-x-y}Ge_ySn_x$ region beneath the a-Si opening. The amorphization depth can be controlled by tiling the sample normal relative to the ion beam. Hence, the implantation in Fig. 7 a) and d) is shallower compared to the implantation parallel to the opening. The presence of Sn in Fig. 7 e) and f) and the still single-crystalline $Si_{0.70}Ge_{0.30}$ layer next to the amorphous $Si_{1-x-y}Ge_ySn_x$ region confirms the successful local Sn implantation.

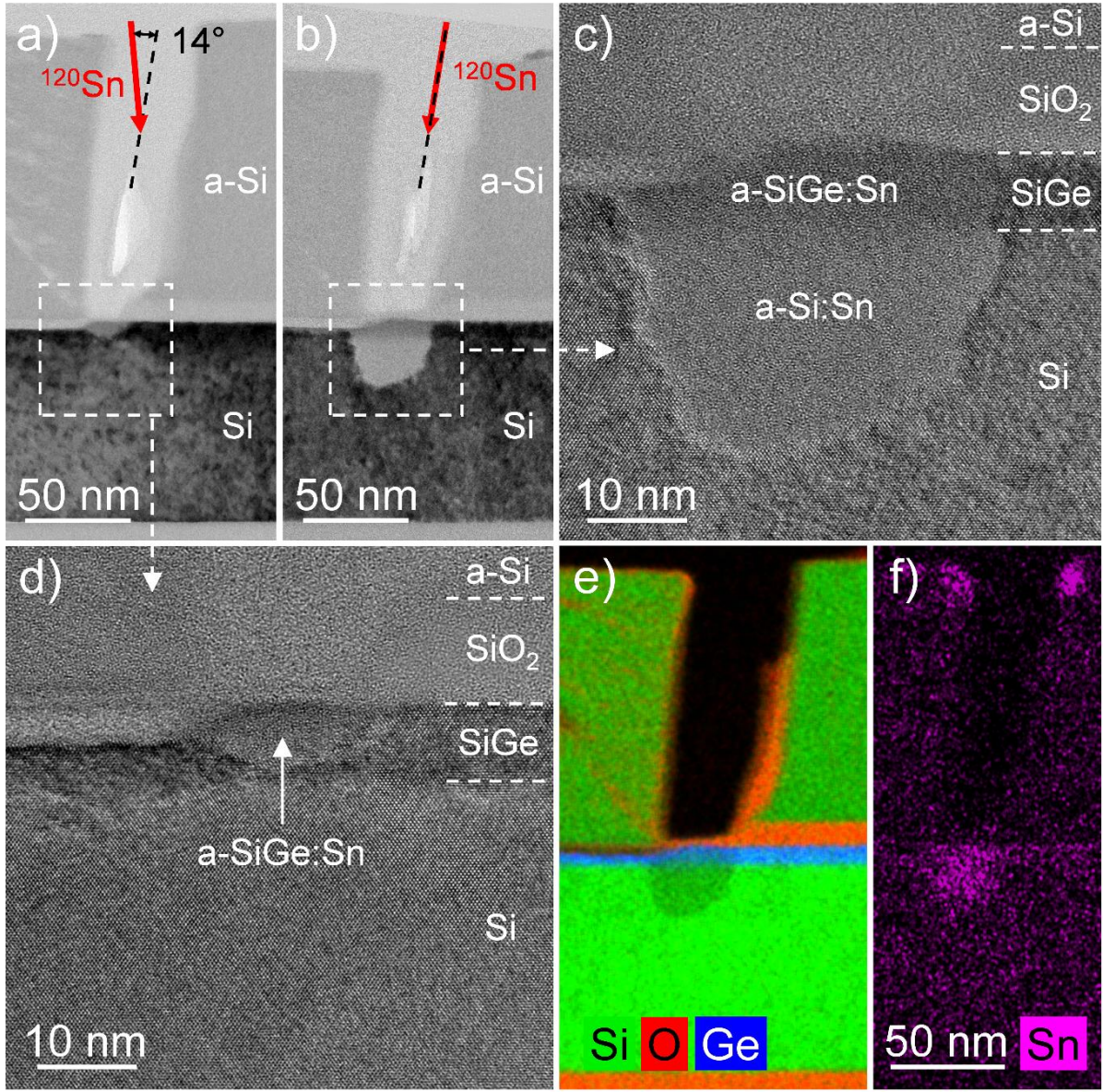

Fig. 7: TEM-based analysis of $Si_{0.70}Ge_{0.30}$ on SOI (sample type C) after implantation of Sn 14° out-of-the cavity axis a) and d), and parallel to the cavity b) and c). The interfaces between the layers are highlighted by white-dashed lines, and the high-resolution TEM images in c) and d) are highlighted by white-dashed squares in panels a) and b), respectively. The EDXS-based element distribution maps in e) and f) were recorded for the region shown in panel b) and represent the superimposed signals from silicon (Si) in green, oxygen (O) in red, and germanium (Ge) in blue e), as well as the corresponding tin (Sn) signal in f).

## IV. Conclusion

The general feasibility of fabricating $Si_{1-x-y}Ge_ySn_x$ alloys by ion beam implantation of Sn into $Si_{1-y}Ge_y$ films and subsequent recrystallization by FLA is shown. In thick Ge-based films, 2.3% Sn could be fully incorporated into the crystal. In the case of Si-rich ultra-thin $Si_{1-y}Ge_y$ films on SOI, good process control for the capping layer deposition and the implantation steps are necessary to avoid too deep or too shallow implantation profiles. About 1% Sn could be partially incorporated into a 15 nm-thick $Si_{0.73}Ge_{0.27}$ layer by r-FLA. The amount of implanted Sn can be further increased by adjusting the implantation angle and achieving deeper implantation profiles. The Sn implantation causes *in situ* strain relaxation during the implantation. It is shown that the $Si_{0.681}Ge_{0.308}Sn_{0.011}$ layer is recrystallized by r-FLA for 20 ms without any indication of Sn segregations. The recrystallized structures after r-FLA are almost free of defects. Furthermore, micro-structuring enables local Sn implantation. The lateral modification of the $Si_{1-x-y}Ge_ySn_x$ alloy composition will locally reduce the band gap after recrystallization. In combination with epitaxial layer growth, this enables lateral and vertical band gap engineering on the same substrate to tune and optimize the properties of electro-optical devices. For example, this approach can pave the way for the fabrication of lateral-heterostructure tunnel field effect transistors.

## V. Supplementary Material

The supplements support the presented article with additional information about the sample structuring process (Supplement A), SRIM simulation results of sample type B and C (Supplement B), and further TEM results of as-implanted $Si_{0.73}Ge_{0.27}$:Sn from sample type B (Supplement C).

## ACKNOWLEDGMENTS

This work was partially supported by the Bundesministerium für Bildung und Forschung (BMBF) under the project "ForMikro": Group-IV heterostructures for high-performance nanoelectronic devices (SiGeSn NanoFETs) (Project-ID: 16ES1075). We gratefully acknowledge the HZDR Ion Beam Center for their support with RBS. The authors thank Romy Aniol and Andreas Worbs for TEM specimen preparation, as well as Dr. rer. nat. Andreas Schubert and Jens Katzer from the Leibniz Institute for High Performance Microelectronics (IHP) for their TEM contributions. Furthermore, the use of the HZDR Ion Beam Center TEM facilities, Blitz-lab, and the funding of TEM Talos by the German BMBF, Grant No. 03SF0451, in the framework of HEMCP are acknowledged. Finally, we acknowledge IHP and Globalfoundries Dresden for the growth of the $Si_{1-y}Ge_y$ layers.

## DATA AVAILABILITY

The data that support the findings of this study are available from the corresponding author upon reasonable request.

# Supplement A: Micro-structuring for local Sn implantation of sample type C

The local implantation was realized by opening a window inside the 115 nm-thick a-Si mask layer. The openings are back-etched spacer structures that were selectively etched by wet chemistry. To fabricate these structures, a ca. 10 nm-thick $SiO_2$ capping layer was deposited (plasma-enhanced CVD at 350 °C with $SiH_4$) first. Afterwards, a 200 nm-thick a-Si layer was added to the entire sample surface by magnetron sputtering (see Fig. A 1 a)).

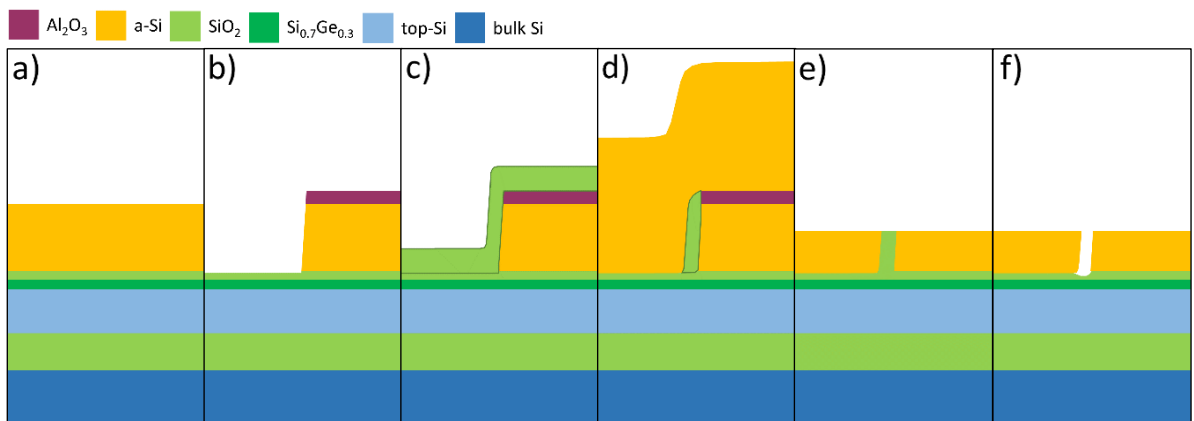

Fig. A 1: Main fabrication steps for the opening. The SiGe on the silicon-on-insulator (SOI) layer stack (see sample type C in Fig. 1) is covered by $SiO_2$ and amorphous Si (a-Si) a). a-Si patterning by using an $Al_2O_3$ hardmask and etching b). Deposition of $SiO_2$ c). Afterwards, $SiO_2$ is thinned down by dry etching and covered by an additional a-Si layer d). Surface polishing down to the $SiO_2$ spacer structure e). Etching of the $SiO_2$ cavity f).

The sample was coated by two PMMA layers with varying molecular weights (AR-P639.04 and AR-P649.04 by supplier Allresist), with the lower-molecular-weight layer being placed first. Subsequent patterning by electron beam lithography and development in methyl isobutyl ketone : isopropyl alcohol (1:3) mixture produced a PMMA layer with undercut at the edges, suitable for a lift-off process. A 40 nm-thick $Al_2O_3$ film was deposited by electron beam evaporation, followed by lift-off in dimethyl sulfoxide at 150 °C. The remaining $Al_2O_3$ structures are very stable in even high-power $SF_6$ dry etch processes and serve as a hard mask to pattern the underlying a-Si layer (Fig. A 1 b)). A precisely tuned process that terminates at the initial 10 nm-thick $SiO_2$ was used. Anisotropic dry etching was conducted with an $SF_6:O_2$ mixture (3:1) at -130 °C to facilitate side wall passivation and achieve a nearly perpendicular a-Si flank. Next, the entire sample was deposited with a 70 nm $SiO_2$ layer by the already mentioned CVD process (see Fig. A 1 c)). Since this step is not fully conformal, roughly half the thickness (≈ 35 nm) is deposited on a perpendicular side wall. Afterwards, a further anisotropic etch selective for $SiO_2$ was employed to precisely thin-down the 70 nm of $SiO_2$ on the sample surface, but not the 35 nm on the side walls. The $CHF_3:Ar$ (1:2) mixture was used at 50 °C to achieve this, resulting in the $SiO_2$ spacer structure. Subsequently, another a-Si deposition of about 230 nm was conducted, which covered the sample and especially the spacer structure from all sides (Fig. A 1 d)). Next, the entire sample surface was polished (mechanically abrasive slurry containing $SiO_2$ nanoparticles) until the spacer structure was revealed at the surface, this time confined between a-Si on both sides (Fig. A 1 e)). The resulting height above the initially 10 nm-thick $SiO_2$ layer was about 115 nm. Lastly, the $SiO_2$ spacer was selectively etched in 1% hydrofluoric acid at room temperature for 10 minutes (see Fig. A 1 f)). The cavity opening has a tilt of about 7° relative to the normal of the sample plane.

## Supplement B: SRIM simulations of the Sn doping profiles

In order to estimate the implantation parameters for the thin-film samples (types B and C), "transport of ions in matter" (TRIM) simulations with the software code "stopping and range of ions in matter" (SRIM) [20] were performed with different $Si_3N_4$ and $SiO_2$ capping layer thicknesses and Sn implantation acceleration energies. At this stage, it must be mentioned that the software SRIM uses some assumptions, like the complete layer stack is amorphous, implantation temperature is 0 K, and every ion is simulated independently from the previously implanted ions (dose always zero). The challenging task is to have, on the one hand, a thick enough capping layer to protect the $Si_{1-y}Ge_y$ surface from surface degradation. On the other hand, the cap should be thin enough to implant the $Si_{1-y}Ge_y$ layer without amorphization of the single-crystalline seed layer.

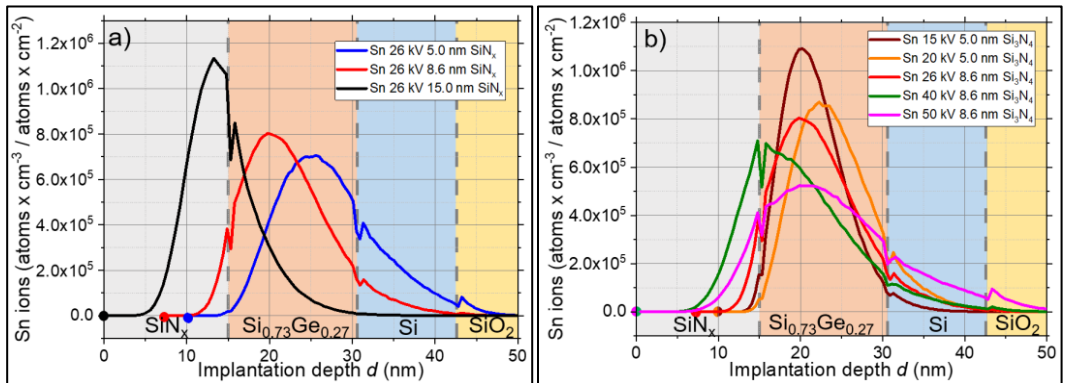

Fig. B 1: a) Simulated Sn implantation depth profile for sample type B with a constant implantation energy of $E$ = 26 kV and different $Si_3N_4$ cap layer thicknesses of 5, 8.6, and 15 nm. b) Simulated Sn implantation depth profile for different implantation energies $E$ = 15, 20, 26, 40, and 50 kV b). The dots correspond to the surface of the $Si_3N_4$ layers.

As visible in Fig. B 1 a) for an acceleration voltage of 26 kV and an 8.6 nm-thick $Si_3N_4$ layer, the maximum Sn concentration is close to the center of the $Si_{0.73}Ge_{0.27}$ layer, and the Sn ion collision cascade stops slightly in front of the BOX $SiO_2$. A reduction of the $Si_3N_4$ layer thickness to 5 nm shifts the peak concentration deeper into the $Si_{0.73}Ge_{0.27}$ layer, leading to an Sn implantation into the $SiO_2$ interface. This deep end-of-range implantation tail can convert the entire seed crystal into an amorphous state (depending on the selected fluence). Hence, amorphization down to the $SiO_2$ must be avoided to maintain a seed crystal. An increase in the $Si_3N_4$ layer thickness shifts the maximum of the Sn peak into the $Si_3N_4$, as visible for 15 nm-thick $Si_3N_4$. This could lead to a shallow Sn implantation close to the $Si_{0.73}Ge_{0.27}$ surface. Another solution would be to adjust the acceleration energy, as shown in Fig. B 1 b). Reducing the ion beam implantation energy from 26 kV to 15 kV for a 5 nm-thick $Si_3N_4$ layer might be a solution in terms of the implantation profile but might be close to the critical threshold thickness to maintain the surface quality. On the other hand, an increase in the implantation energy to 50 kV for 15nm $Si_3N_4$ would cause a long tail of collision cascades and can destroy the Si seed crystal. Therefore, an $Si_3N_4$ capping layer below 10 nm is favorable for this $Si_{1-x-y}Ge_ySn_x$ fabrication approach of sample type B.

For sample type C, different $SiO_2$ capping layer thicknesses were simulated, as shown in Fig. B 2, using the approach explained above. Based on the simulation results, the 10.4 nm-thick $SiO_2$ was selected for the experiments.

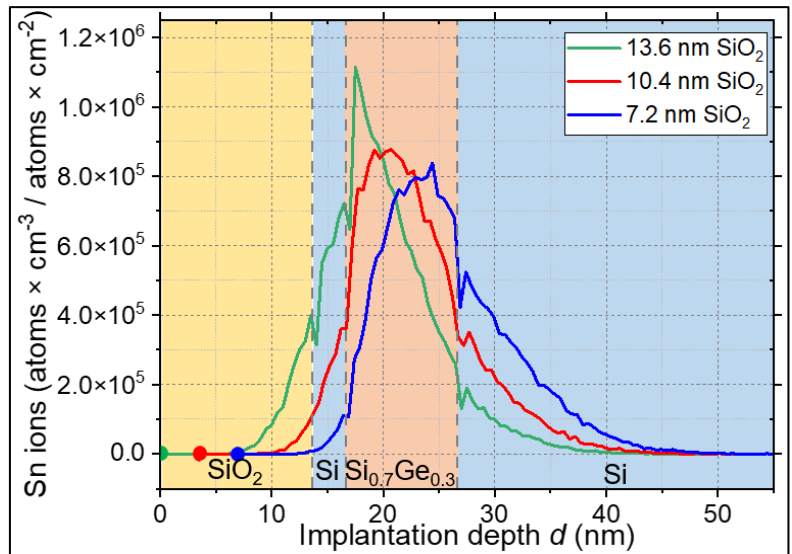

Fig. B 2: Simulated Sn implantation depth profile for sample type C with a constant implantation energy of $E$ = 24 kV and different $SiO_2$ layer thicknesses of 7.2, 10.4, and 13.6 nm. The dots correspond to the surface of the $SiO_2$ layer.

# Supplement C: TEM characterization of as-implanted $Si_{0.73}Ge_{0.27}$:Sn of sample type B

Fig. C 1 shows recurring amorphous $Si_{1-x-y}Ge_ySn_x$ regions (light-gray contrast) within the crystalline $Si_{1-x-y}Ge_ySn_x$ matrix layer (dark contrast). In the cross-section, the amorphous fractions seem to have a triangular shape or are wavelike distributed across the sample surface.

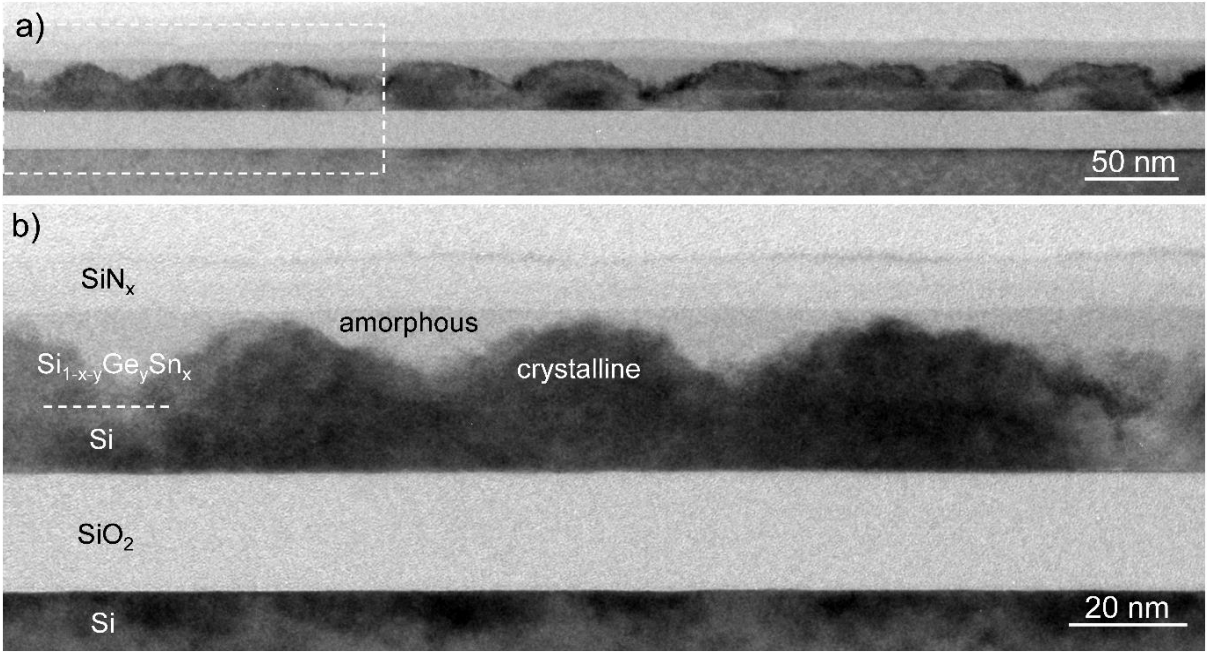

Fig. C 1: Corss-sectional bright-field TEM images of the sample type B after Sn implantation. Panel b) shows the magnified field of view indicated by the white-dashed rectangle in a).